\documentclass[a4paper,dvips 2t,nofootinbib,notitlepage]{revtex4-1}
\usepackage{epsfig,graphics,graphicx,amsmath,amssymb}

\begin{document}

\title{\bf Analyzing of singlet fermionic dark matter via the updated direct detection data }

 \author{ M.M. Ettefaghi}\email{mettefaghi@qom.ac.ir} \author{ R. Moazzemi}\email{r.moazzemi@qom.ac.ir}

\affiliation{Department of Physics, University of Qom, Ghadir Blvd., Qom 371614-611, I.R. Iran}

\begin{abstract}

We revisit the parameter space of singlet fermionic cold dark matter model in order to determine the role of the mixing angle between the standard model Higgs and new singlet one. Furthermore, we restudy the direct detection constraints with the updated and new experimental data. As an important conclusion, this model is completely excluded by recent  XENON100, PandaX II and LUX  data.

\end{abstract}
\maketitle

\section{Introduction}

There exist several pieces of evidence that indicate the highest fraction of matter in the universe is composed of unknown
particles called dark matter (DM) (see \cite{rev1, rev2}). The baryonic matter composes
only less than 5\% of the universe content. While the standard model (SM) is very successful in the experimental tests, it does not predict any appropriate candidate for DM. Hence, many authors have been convinced  that we need a model beyond the SM. The evidence hints that the DM candidates should be mostly stable,
non-baryonic, massive, non-relativistic and have insignificant or very weak interactions with
other particles (see \cite{Taoso} for a discussion of the conditions of DM candidates and their properties).
The DM particles with these properties are often called cold DM (CDM) or weakly interacting
massive particles (WIMPs). Since no signal, predicted by any theory beyond SM, has been confirmed experimentally, it is reasonable to consider the most minimal extension of the SM to explain DM. Singlet fermionic CDM (SFCDM) is a minimal extension of the SM which proposes a singlet fermion as an appropriate candidate for CDM \cite{ kim1,kim2,Beak,ettefaghi}.

One can achieve a renormalizable theory for SFCDM if the SM is extended by a
singlet fermion as CDM and a singlet scalar Higgs boson as a mediator. For the SFCDM masses below 100 GeV,  the relic abundance constraint and the direct detection bounds have been studied in \cite{kim2,Beak}. An almost comprehensive study of the parameter space of SFCDM has been performed in \cite{ettefaghi}. The
SFCDM annihilation into two photons under the relic abundance constraint has been obtained
and compared with Fermi-Lat bounds for masses below 200 GeV in \cite{ettefaghi2}. From the Higgs coupling measurements, the mixing
angle is constrained at 95\% CL to be $\sin\theta\lesssim 0.4$ \cite{atlastheta}, independent of the second Higgs mass. The analysis of Ref. \cite{falkow}, by the electroweak
precision tests, implies slightly stronger constraints in the relevant mass range; for example one finds $\sin\theta\leq 0.32$ for  the second Higgs mass about  750 GeV  at 95\% CL. In addition, for this mass of the Higgs, it has been shown that $\sin\theta$ is constrained to be less than 0.1, and this constraint is also put on any scenario where the new scalar is somehow involved in electroweak symmetry breaking \cite{adam}. In this paper, we restudy the parameter space of the SFCDM, focusing on the role of the Higgs mixing angle and compare our results with latest experimental data. We take the SM and singlet Higgs mass to be 125 and 750 GeV, respectively. The former is fixed by earlier ATLAS \cite{atlas} and CMS \cite{cms} results. For the latter, due to the above statements for the Higgs mixing angle, we choose 750 TeV as an interesting mass.\footnote{It is also notable that an excess in the diphotons events with the invariant mass of about 750 GeV has been reported by ATLAS \cite{atlas2015} and CMS \cite{cms2015} based on data collected in 2015, though the analyzes based on data collected in 2016 \cite{atlas2016,cms2016} show no significant excess over the SM predictions.}Of course, as we shall state in Sect. \ref{sec3a}, for the other masses between the range about 500-1000 GeV our general results and discussions do not get altered.

Furthermore, there are several experiments which report the measured cross section for direct detection  of dark matter, recently, such as the XENON100, LUX, COUPP, PICO,  EDELWIESS II, PandaX II and Darkside Collaborations. In this paper, using the most updated direct detection data reported by some of these experiments and considering the issues on the mixing angle mentioned above, we reanalyze the parameter space by imposing the relic abundance condition. We shall see that the entire parameter space is excluded by XENON100 \cite{XENON100}, PandaX II \cite{pandax}, and LUX \cite{LUX2016}.

We have organized the paper as follows: In Sect. \ref{model} the renormalizable model for a SFCDM is briefly reviewed. In Sect. \ref{sec3}, we obtain the coupling constant by imposing the relic abundance condition, then we calculate the scattering cross section of SFCDM from the nucleon and explore the parameter space using the most recent direct detection data. Finally, we summarize our discussion and conclusions in the last section.

\section{The model}\label{model}

The most minimal extension of the SM, including a CDM candidate, is achieved by adding a gauge singlet fermion.  We can consider the singlet fermion to play the dark matter role (SFCDM) provided that it has a very weak interaction with the SM particles because it must respect the relic abundance condition. To accommodate this in a renormalizable manner, a singlet Higgs $S$, in addition to the usual Higgs doublet, is needed as mediator between SFCDM and the SM particles \cite{kim2,ettefaghi}.
The Lagrangian for the SFCDM model can be decomposed as follows:
\begin{equation} \label{sfcdm}
{\cal L}_{\text{SFCDM}}={\cal L}_{\text{SM}}+{\cal L}_{\text{hid}}+{\cal L}_{\text{int}},
\end{equation}
where ${\cal L}_{\rm{SM}}$ is the SM Lagrangian and ${\cal L}_{\rm{hid}}$ denotes  the hidden sector Lagrangian,
\begin{eqnarray}\label{lhid}
 {\cal L}_{\text{hid}}={\cal L}_\psi+{\cal L}_{\text{S}}-{g_s \overline \psi  \psi S}.
\end{eqnarray}
Here, ${\cal L}_\psi$ and ${\cal L}_{\text{S}}$ are the free Lagrangians of SFCDM,
\begin{equation}{\cal L}_\psi=\bar{\psi}(i\partial\!\!\!/-m_{\psi_0})\psi,\end{equation}
and the singlet Higgs,
\begin{equation}\label{selfS}{\cal L}_{\text{S}}=\frac 1 2 (\partial_\mu S)(\partial^\mu S)-\frac{m_0^2}{2}S^2-\frac{\lambda_3}{3!}S^3-\frac{\lambda_4}{4!}S^4.\end{equation}
The last term in Eq. (\ref{lhid}) is due to the interaction between the SFCDM and singlet Higgs with coupling constant $g_s$. In Eq. (\ref{sfcdm}), ${\cal L}_{\text{int}}$ is related to the interaction between the new singlet Higgs and the SM doublet one
\begin{equation}  \label{hs}
{\cal L}_{\text{int}}=-\lambda_1H^\dag HS-\lambda_2H^\dag
HS^2.
\end{equation}
We have $ \langle H\rangle=\frac{1}{\sqrt{2}}\left(
                             \begin{array}{c}
                               0 \\
                               v_0
                             \end{array}\right)$
and $\langle S\rangle=x_0$, with $v_0$ and $x_0$ being the vacuum expectation values (VEV) of the SM Higgs and singlet Higgs, respectively.
We define the fields $h$ and $s$ as the fluctuation around the VEVs of them. Therefore, after symmetry breaking we have
\begin{equation}
 H=\frac{1}{\sqrt{2}}\left(
                             \begin{array}{c}
                               0 \\
                               h+v_0
                             \end{array}\right),
\end{equation}
and
\begin{equation}
S=s+x_0.
\end{equation}
We can obtain the mass eigenstates by diagonalizing the mass matrix as follows:
\begin{eqnarray}
 h_1=\sin\theta s+\cos\theta h,\nonumber \\
h_2=\cos\theta s-\sin\theta h,
\end{eqnarray}
where  $\theta$ is a mixing angle which depends on the parameters of the Lagrangian (\ref{sfcdm}).
One naturally expects that $|\cos\theta|>\frac{1}{2}$, so that $h_1$ is the SM Higgs-like scalar, while $h_2$ is the singlet-like one.
The singlet fermion  has mass $m_\psi=m_{\psi_0}+ g_Sx_0$, which is an independent parameter in the model. The VEV of our singlet Higgs, $x_0$, is completely determined by minimization of the total potential (including SM and singlet Higgs potentials) as follows:
 \[x_0=-\frac{1}{4v\lambda_2}\left[(m_{h_1}^2+m_{h_2}^2-4v^2\lambda_0)\tan2\theta+2v\lambda_1\right],\]
 where $\lambda_0$ is the self-interaction coupling constant of the SM Higgs.
 There are seven independent parameters, in addition to the SM ones, in this model: $\{m_\psi,g_S,m_0,\lambda_1,\lambda_2,\lambda_3,\lambda_4\}$.  After spontaneous symmetry breaking we encounter a new set of parameters,  $m_\psi$, $g_S$, second Higgs mass $m_{h_2}$, $\lambda_1$, $\lambda_2$, $\lambda_3$, $\lambda_4$ and  the mixing angle between Higgs bosons $\theta$, which is not an independent parameter.

\subsection{The cross section}
\begin{figure}[th]
\centerline{\includegraphics[width=15.5cm]{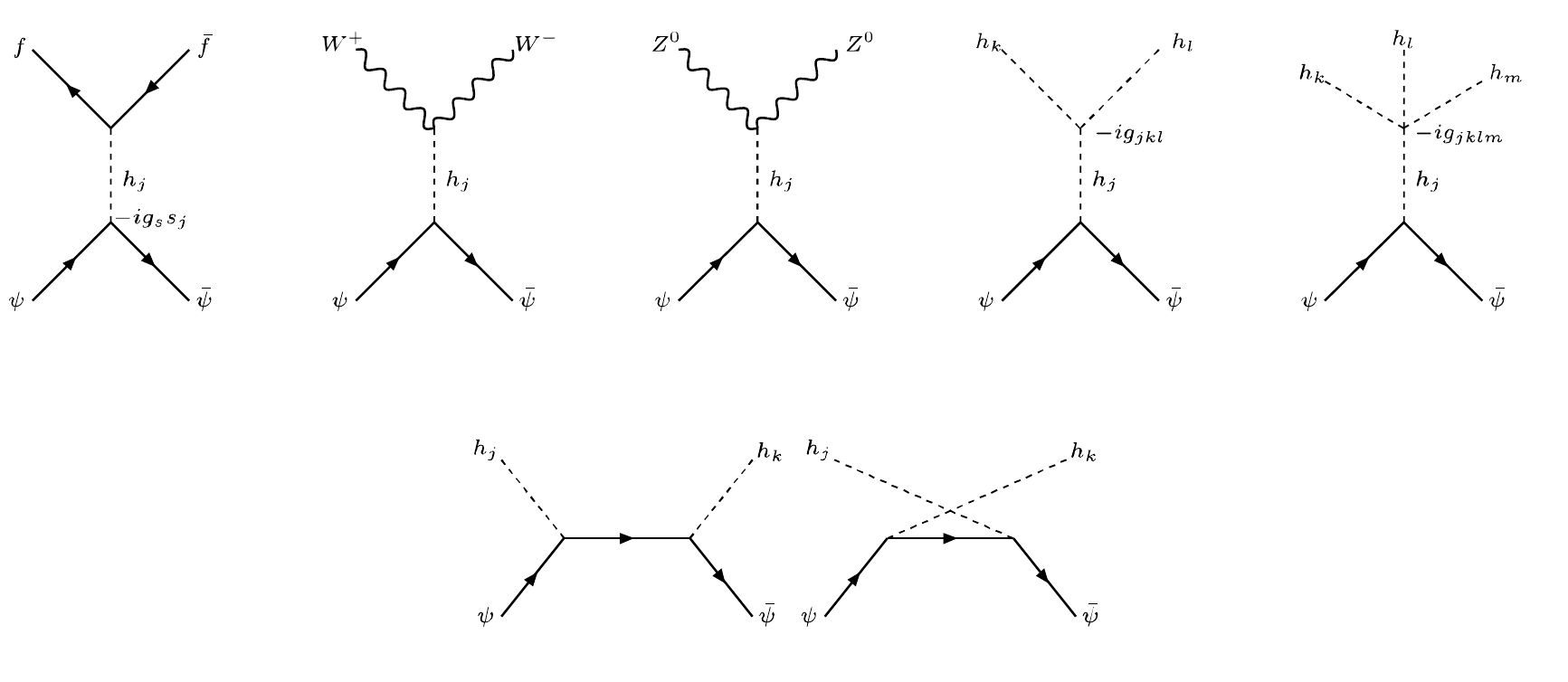}}\caption{The Feynman diagrams for the annihilation of  singlet fermion pairs into SM particles, two and three Higgs bosons at tree level. The vertex factor of three (four) Higgs boson lines,  $-ig_{ijk}$ ($-ig_{ijkl}$), is symmetric under permutations of their subscripts. For three Higgs bosons in final state only the dominant Feynman diagrams are shown. Obviously, the first row is due to the $s$-channel while the second row indicates the $t$- and $u$-channels.}\label{fig11.}
\end{figure}

In the SFCDM  model, at tree level, pairs of singlet fermions can annihilate into SM particles, including pairs of massive fermions and gauge bosons, and also two and three Higgs bosons. We have listed the corresponding Feynman diagrams in Fig. \ref{fig11.}. These diagrams are at leading order, so we should respect the perturbation criteria in our calculations.

We have calculated the corresponding cross section of this annihilation process. According to Fig.  \ref{fig11.}, while the annihilations into the fermions and gauge bosons proceed  only through the $s$-channel, the annihilation into Higgs bosons occurs via the $s$-, $t$- and $u$-channels. The total annihilation cross section times the relative velocity $v$ can be written as follows:
\begin{equation}\label{sigmaa}
 { {\sigma v}}_{\text{ann}}  = {\sigma v }_{\text{SM}}+{\sigma v }_{\text{2Higgs}}+ {\sigma v}_{\text{3Higgs}},
\end{equation}
where the $ { {\sigma v}}_{\text{SM}} $ is given by

\begin{eqnarray}
\sigma v_{\text{SM}} &=& \frac{(g_s s_1 s_2)^2}
                        {16\pi}
                   \left(1-\frac{4 m_{\psi}^2}{{s}}\right)
\nonumber\\
      & & \times \left(\sum_{j=1,2}\frac{1}
            {d_j}
            -\frac{2({s}-m_{h_1}^2)({s}-m_{h_2}^2)
                    + 2 m_{h_1}m_{h_2}\Gamma_{h_1}\Gamma_{h_2}}
                  {d_1d_2}
      \right)
\nonumber\\
      & & \times \left[\sum_{f(\text{fermions})}
           2\lambda_fs\left(\frac{m_f}{v_0} \right)^2 \left(1-\frac{4 m_f^2}{{s}}\right)^{3/2}
      \right.
\nonumber\\
      & & ~~~~~
         \left.  + \sum_{{w=W^{^+},W^{^-},Z^0}}2\left(\frac{m_w^2}{v_0}\right)^2
             \left(2+\frac{({s}-2m_w^2)^2}{4 m_w^4}\right)
                 \sqrt{1-\frac{4 m_w^2}{{s}}}\right],
\end{eqnarray}
where $\lambda_f$ is 3 (1) for quarks (leptons), $\Gamma_{h_j}$ refers to the decay widths of $h_j$ and $d_j=({s}-m_{h_j}^2)^2 + m_{h_j}^2 \Gamma_{h_j}^2$ ($j=1,2$). Here, we have used the abbreviations $s_1\equiv\sin\theta$ and $s_2\equiv\cos\theta$. The last two terms in Eq. (\ref{sigmaa}) are  the annihilation cross sections into two and three Higgs bosons, respectively. To obtain these cross sections we should derive $g_{jkl}$ and $g_{jklm}$ corresponding to the vertex factors of them. For $j\neq k$ we get
\begin{eqnarray}\label{couplings}
g_{{jjj}}&=&\frac{1}{3} \left\{6 (-1)^j v_0 s_k \left(\lambda _2 s_j^2+\lambda _0 s_k^2\right)-s_j \left[s_j^2 \left(\lambda _3+\lambda _4 x_0\right)+3 \lambda _1 s_k^2\right]\right\},
   \nonumber\\
g_{{jkk}}&=&\frac{1}{4} \left\{2 (-1)^k v_0 s_j \left[\lambda _2 \left(1-3 s_j^2\right)+3 \left(4 \lambda _0-3 \lambda _2\right) s_k^2\right]
\right.,
 \nonumber\\
&& \qquad\left.
+s_j^2 s_k \left[9 \lambda _1-4 \lambda _3+2 \left(9 \lambda _2-2 \lambda _4\right) x_0\right]-\left(3 s_k^3+s_k\right) \left(\lambda _1+2 \lambda _2 x_0\right)\right\},
\nonumber\\
g_{jjjj}&=&-12 \lambda _2 s_1^2s_2^2-\lambda _4 s_j^4-6 \lambda_0 s_k^4,
 \nonumber\\
 g_{1122}&=&\frac{1}{8}\left\{[\cos (4 \theta )-1] \left(\lambda _4+6 \lambda_0\right)-4 \lambda _2 [3 \cos (4 \theta )+1]\right\},
\nonumber\\
g_{jjjk}&=&s_2 s_1\left(6 \lambda _2 (s_j^2-s_k^2)-\lambda _4 s_j^2+6 \lambda_0 s_k^2\right).
\end{eqnarray}
Note that $g_{jkl}$ and $g_{jklm}$ are symmetric under permutation of their subscripts and $j,k,l,m=1,2$. Therefore, one can derive the annihilation cross section into two Higgs bosons as follows:
\begin{eqnarray}
\sigma v_{\text{2Higgs}}&=&\frac{g_s^2}{16 \pi } \left(1-\frac{4 m_{\psi }^2}{s}\right) \left\{-\frac{4 g_s^2 s_1^2 s_2^2}{y \left(y^2-1\right) \left(-m_{h_1}^2-m_{h_2}^2+s\right){}^2}
\right.
\nonumber\\&\times&
\left.
 \left\{\left(-m_{h_1}^2-m_{h_2}^2+s\right){}^2 y^3+\left[-32 m_{\psi }^4+8 \left(m_{h_1}^2+m_{h_2}^2\right) m_{\psi }^2-m_{h_1}^4-\left(m_{h_2}^2-s\right){}^2-m_{h_1}^2 \left(4 m_{h_2}^2-2 s\right)\right] y
\right.\right.
\nonumber\\&&
\left.\left.
+\left(y^2-1\right) \tanh ^{-1}y \left[32 m_{\psi }^4+8 \left(m_{h_1}^2+m_{h_2}^2-2 s\right) m_{\psi }^2-m_{h_1}^4-\left(m_{h_2}^2-s\right){}^2+2 m_{h_1}^2 \left(s-2 m_{h_2}^2\right)\right]\right\}
\right.
\nonumber\\&-&
\left.
\frac{8g_s m_{\psi } s_1 s_2}{d_1 d_2} \left[\frac{\tanh ^{-1}y \left(8 m_{\psi }^2-m_{h_1}^2-m_{h_2}^2-s\right)}{y \left(-m_{h_1}^2-m_{h_2}^2+s\right)}-1\right] \left[d_2 g_{112} \left(s-m_{h_1}^2\right) s_1+d_1 g_{212} \left(s-m_{h_2}^2\right) s_2\right]
\right.
\nonumber\\&+&
\left.
\sqrt{{\frac{\left(-m_{h_1}^2-m_{h_2}^2+s\right){}^2-4 m_{h_1}^2 m_{h_2}^2}{s^2}}} \left[\frac{2 g_{112} g_{212} s_1 s_2}{d_1 d_2} \left[\left(s-m_{h_1}^2\right) \left(s-m_{h_2}^2\right)+m_{h_1} m_{h_2} \Gamma _{h_1} \Gamma _{h_2}\right]+\sum _{j=1,2} \frac{g_{{j12}}^2 s_j}{d_j}\right]
\right.
\nonumber\\&+&
\left.
\frac{1}{2} \sum _{k=1,2} \left[\frac{g_s^2  s_k^4}{ x_k \left(x_k^2-1\right)\left(s-2 m_{h_k}^2\right)^2} \left[4 x_k \left(32 m_{\psi }^4-16 m_{h_k}^2 m_{\psi }^2+6 m_{h_k}^4+s^2-4 s m_{h_k}^2-\left(s-2 m_{h_k}^2\right){}^2 x_k^2\right)
\right.\right.\right.
\nonumber\\&&
\left.\left.\left.
-4\left(x_k^2-1\right) \tanh ^{-1}x_k \left(32 m_{\psi }^4+16 \left(m_{h_k}^2-s\right) m_{\psi }^2-6 m_{h_k}^4-s^2+4 s m_{h_k}^2\right) \right]
\right.\right.
\nonumber\\&&
\left.\left.
+\sqrt{1-\frac{4 m_{h_k}^2}{s}} \left(\frac{2 g_{{1kk}} g_{2 {kk}} s_1 s_2}{d_1 d_2}{ \left[\left(s-m_{h_1}^2\right) \left(s-m_{h_2}^2\right)+m_{h_1} m_{h_2} \Gamma _{h_1} \Gamma _{h_2}\right]}+\sum _{j=1,2} \frac{g_{{jkk}}^2 s_j}{d_j}\right)
\right.\right.
\nonumber\\&&
\left.\left.
-\frac{8 \left({g_s} m_{\psi } s_k^2\right)}{d_1 d_2}{ \left(\frac{\tanh ^{-1}x_k\left(-8 m_{\psi }^2+2 m_{h_k}^2+s\right)}{\left(2 m_{h_k}^2-s\right) x_k}-1\right) \sum _{j=1,2} g_{{jkk}} \left(s-m_{h_j}^2\right) s_j d_j}\right]\right\},
\end{eqnarray}
where
\[x_k={\sqrt{1-\frac{4 m_{\psi }^2}{s}} \sqrt{1-\frac{4 m_{h_k}^2}{s}}}\bigg/{\left(1-\frac{2 m_{h_k}^2}{s}\right)},\]
 and
 \[y=-{\sqrt{1-\frac{4 m_{\psi}^2}{s}} \sqrt{\frac{m_{h_1}^4}{s^2}+\left(\frac{m_{h_2}^2}{s}+1\right) \left(1-\frac{2 m_{h_1}^2+m_{h_2}^2}{s}\right)}}\bigg/{\left(1-\frac{m_{h_1}^2+m_{h_2}^2}{s}\right)}.\]
Although the annihilation cross section into three Higgs bosons is suppressed due to its narrow phase space integral, to have a complete and more precise  calculation we take it into account. For this term we have
\begin{eqnarray}
\sigma v_{\text{3Higgs}} =\frac{2g_{s}^{2}(s-4 m_{\psi}^{2})}{1536\pi^{3}}\sum _{k,l,m}\left(\sum _{j=1,2} \frac{g_{{jklm}}^2 s_j^2}{d_j}+\frac{2 s_1 s_2 g_{1{klm}} g_{2{klm}} \left[\Gamma _{h_1} \Gamma _{h_2} m_{h_1} m_{h_2}+\left(s-m_{h_1}^2\right) \left(s-m_{h_2}^2\right)\right]}{d_1 d_2} \right).
\end{eqnarray}

\section{Computations}\label{sec3}
\subsection{The relic density}\label{sec3a}
The relic density $\Omega_\psi h^2$, defined as the ratio of the present density of particles to the critical density, is written as follows:
\begin{equation}
\Omega_\psi h^2\approx\frac{(1.07\times10^9)x_F}{\sqrt{g_*}M_{\text{Pl}}(GeV)\left\langle\sigma v_{\text{ann}}\right\rangle},
\end{equation}
where $\left\langle\sigma v_{\text{ann}}\right\rangle$ is the thermally averaged annihilation cross section times the relative velocity \cite{gondolo}:
\begin{equation}
\left\langle\sigma v_{\text{ann}}\right\rangle=\frac{1}{8m_\psi^4 T_FK_2^2\left(\frac{m_\psi}{T_F}\right)}\int_{4m_\psi^2}^\infty ds\sigma _{\text{ann}} \left( s \right)\left(s-4m_\psi^2\right)\sqrt{s}K_1\left(\frac{\sqrt{s}}{T_F}\right),
\end{equation}
with $K_{1,2}(x)$ being the modified Bessel functions. Here $x_F=m_\psi/T_F$ is the inverse freeze-out temperature, which can be determined by the following iterative equation:
\begin{equation}
x_F=\ln\left(\frac{m_\psi}{2\pi^3}\sqrt{\frac{45M_{\text{Pl}}}{2g_*x_F}}\left\langle\sigma v_{\text{ann}}\right\rangle\right),
\end{equation}
\begin{figure}[th]
	\centerline{\includegraphics[width=16cm]{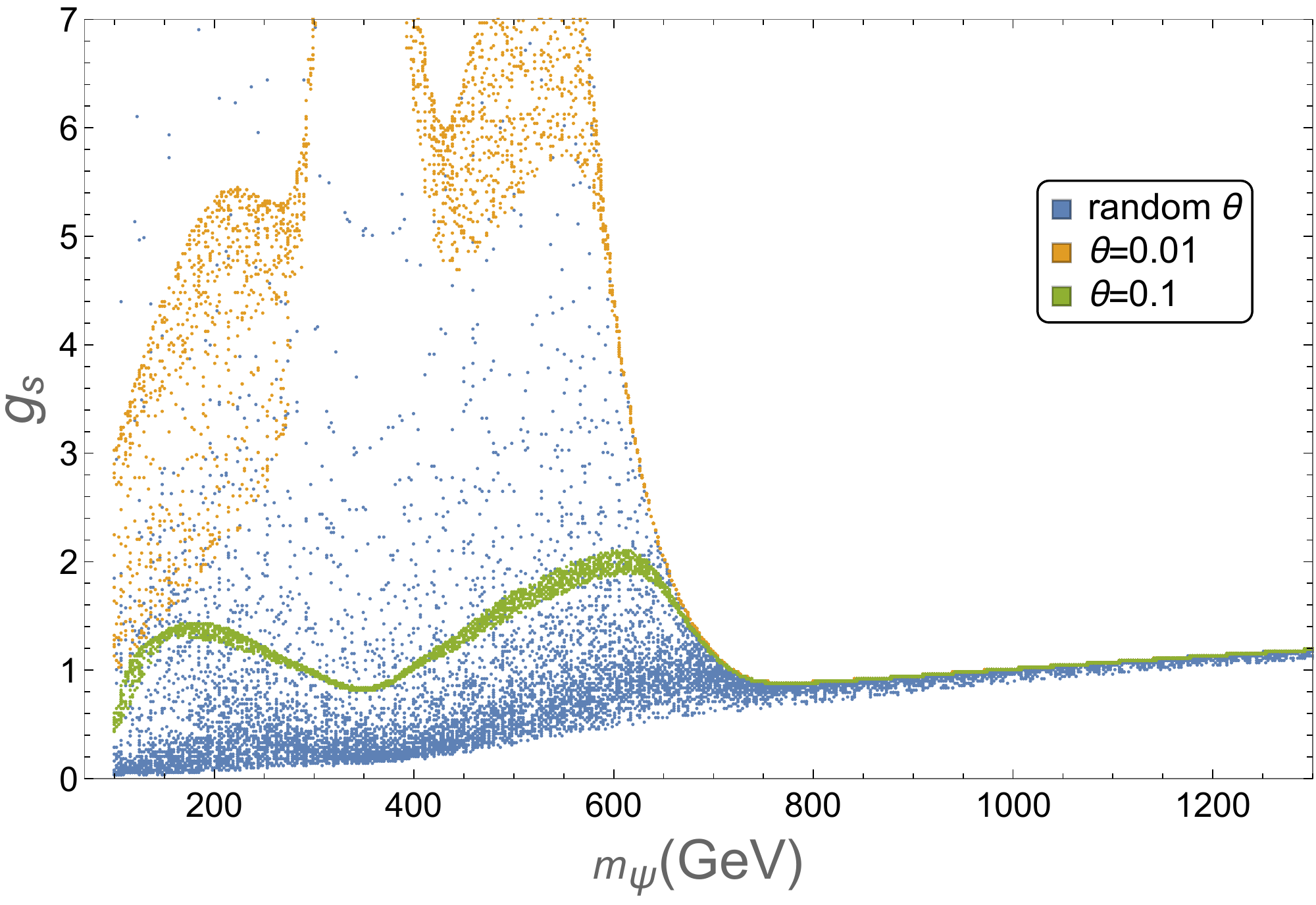}}\caption{The coupling $g_s$ in terms of the dark matter mass $m_\psi$ for three different types of choice of the mixing angle $\theta$.}\label{gs}
\end{figure}
\begin{figure}[th]
	\centerline{\includegraphics[width=10cm]{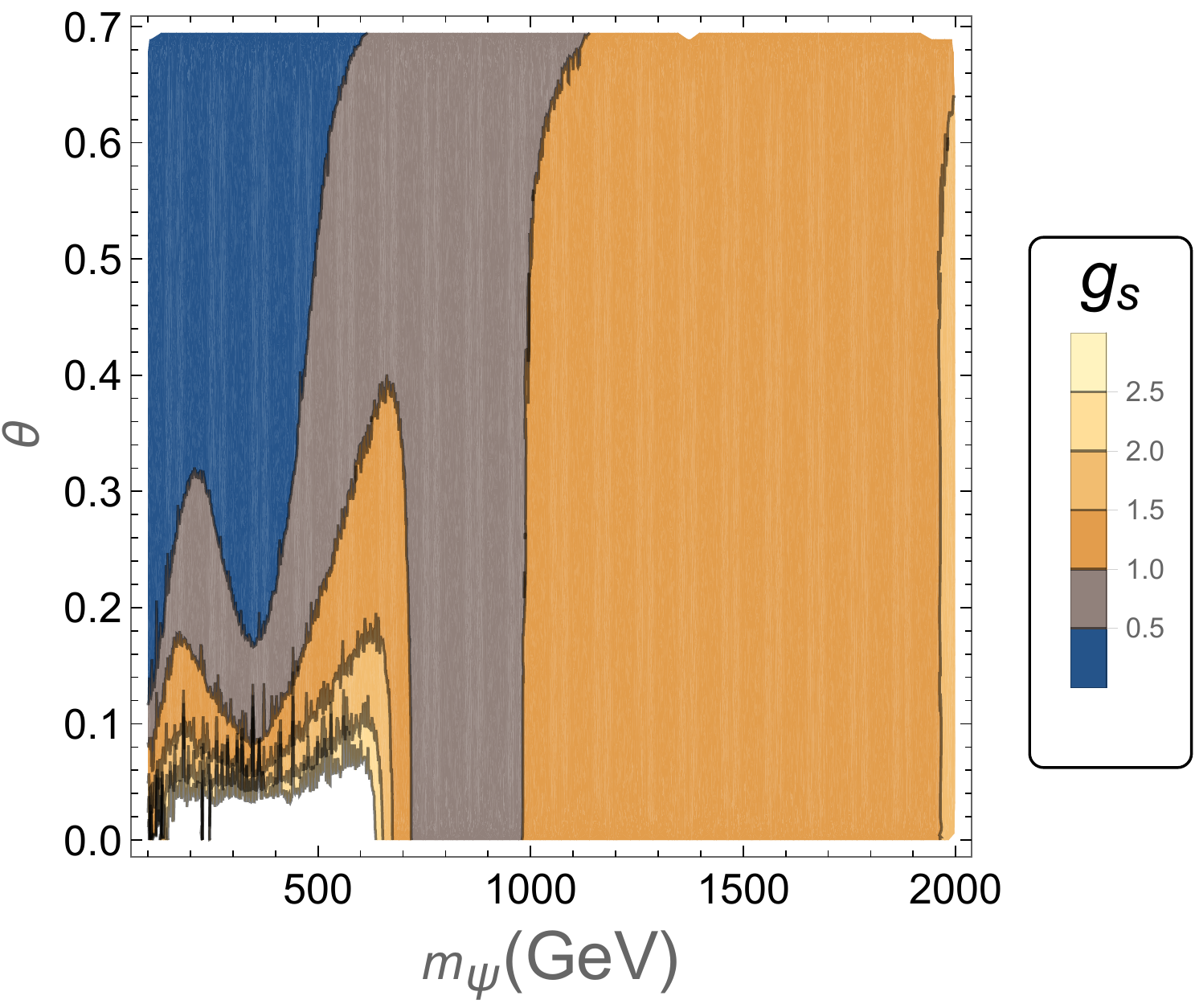}}\caption{The contour plot of the $\theta$ in terms of dark matter mass $m_\psi$; the color illustrates the change in coupling $g_s$.}\label{theta}
\end{figure}
where $g_*$ is the effective degrees of freedom for the relativistic quantities in equilibrium \cite{colb} and $M_{\rm{Pl}}=1.22\times 10^{19}$GeV is the Planck mass.

To study the allowed parameter space consistent with the relic abundance constraint obtained by WMAP observations \cite{Spergel}, the SM Higgs boson mass is fixed to 125 GeV according to the 2012 CMS and ATLAS results \cite{atlas,cms} and the other Higgs mass to 750 GeV.\footnote{From Ref. \cite{ettefaghi} one can see, for wide range masses (about 500-1000 TeV) the minimum of $g_s$ does not change seriously, so that  the direct detection cross section cannot fall below the XENON100 bound (see Fig. \ref{fig1.}).} Although the variations of the $\lambda$'s have no significant impact \cite{ettefaghi}, we let them vary as far as  perturbation theory is correct.
To find the couplings $g_s$ which satisfy the relic density condition, we first investigate about 25000 sample models randomly in the whole parameter space. Namely, in addition to $\lambda$'s, we take $\theta$ and $m_\psi$ to be free. In the other two investigations, each of which concerned whit 10000 sample models, we set $\theta=0.1$ and $\theta=0.01$. We collect all of these three data sets in Fig. \ref{gs}. Using our first data set, we also illustrate the role of the mixing angle $\theta$ by the contour plot of Fig. \ref{theta}. This figure shows that for $\theta<0.1$ there is only a mass region between about 700-1000 GeV as well as a narrow one about 350 GeV, where we get $g_s<1$ and therefore our perturbative analysis works self-consistently. For the other regions, although  obtaining  $g_s$ from the relic density is not consistent with perturbation theory, we necessarily conclude that $g_s>1$.

\subsection{Direct detection}
In this subsection, we investigate the consistency of SFCDM with the direct detection bounds.
 We use the following effective Lagrangian at the hadronic level to describe the scattering of SFCDM from a nucleon:
\begin{equation}
\label{eff}
  {\cal L}_{\text{eff}} =f_{\text{p}}(\bar{\psi}\psi)(\bar{\text{p}}\text{p})+f_{\text{n}}(\bar{\psi}\psi)(\bar{\text{n}}\text{n}),
\end{equation}
where $f_{\text{p}}$ and $f_{\text{n}}$  are the effective couplings of DM to protons and neutrons, respectively, and they are given by:
\begin{equation}\label{6}
 \frac{ f_{\text{p,n}}}{m_{\text{p,n}}}=\sum_{q=u,d,s} f_{Tq}^{\text{(p,n)}}\frac{\alpha_{q}}{m_{q}}+\frac{2}{27}f_{Tg}^{\text{(p,n)}}\sum_{q=c,b,t} \frac{\alpha_{q}}{m_{q}},
\end{equation}
with the matrix elements $m_{\text{p,n}}f_{Tq}^{\text{(p,n)}}\equiv \langle \text{p,n}|m_{q}\bar{q}q|\text{p,n}\rangle$  for $q=u,d,s$ and $f_{Tg}^{\text{(p,n)}} =1-\sum\limits_{q=u,d,s} f_{Tq}^{\text{(p,n)}}$. The numerical values of the hadronic matrix elements are given in \cite{Ellis}. Here, $\alpha_q$ is an effective coupling constant between SFCDM and quark $q$, in the following effective Lagrangian:
\begin{equation}
{\cal L}_{\text{eff}}=\sum_q\alpha_q\bar{\psi}\psi\bar{q}q.
\end{equation}
Since the scattering SFCDM and quarks proceeds through $t$-channel by intermediating a Higgs boson, $\alpha_q$ can be derived:
\begin{equation}
\alpha_q=\frac{g_s \sin\theta\cos\theta m_q}{v_0}\left(\frac{1}{m_{h_1}^2}-\frac{1}{m_{h_2}^2}\right).
\end{equation}
Consequently, the elastic spin-independent scattering cross section off a single nucleon becomes
\begin{equation}\label{j}
 \sigma(\psi \text{p}\rightarrow \psi \text{p})= \frac{4m_\text{r}^{2}}{\pi}f_{\text{p}}^{2},
\end{equation}
\begin{figure}[th]
	\centerline{\includegraphics[width=18cm]{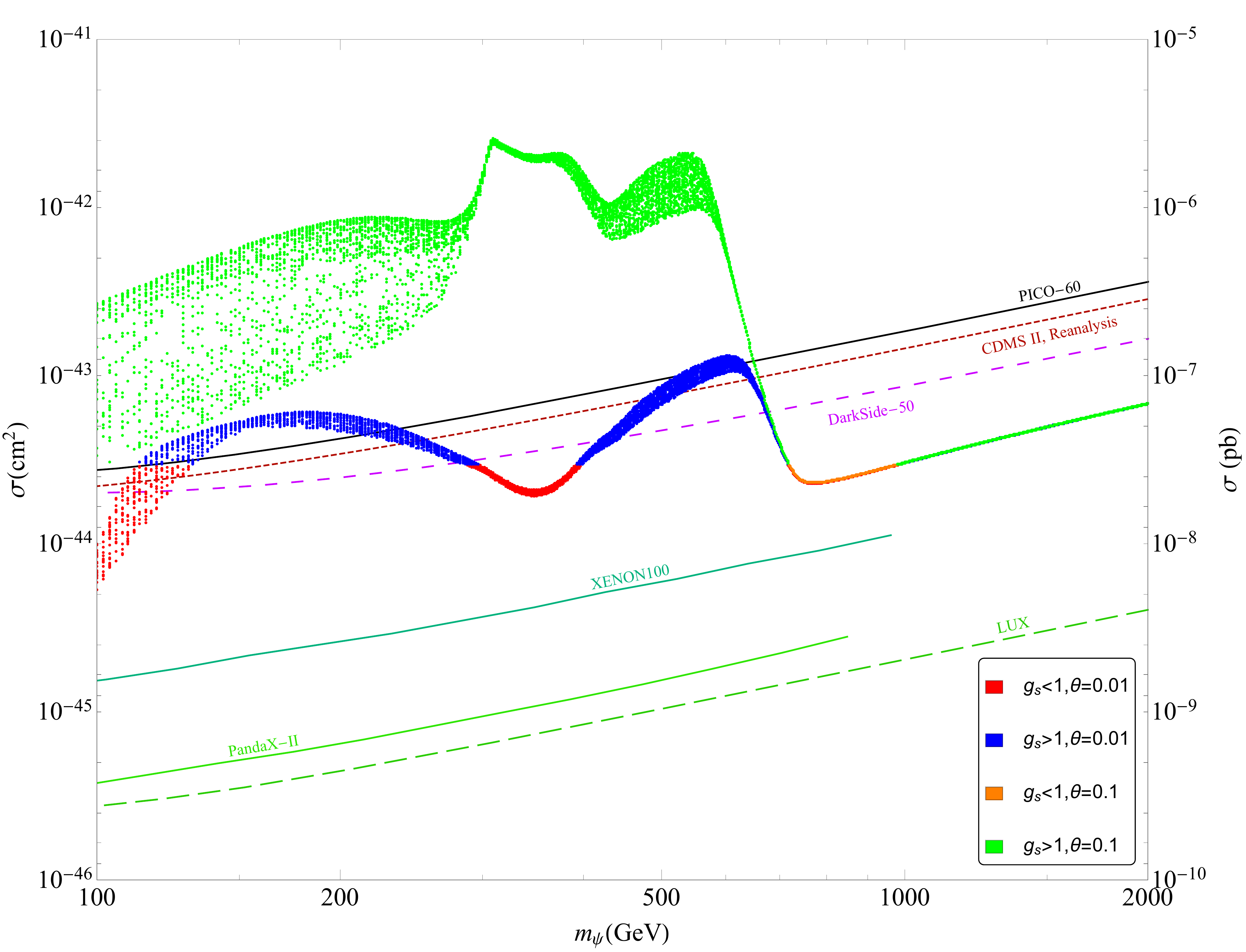}}\caption{The elastic scattering cross section off a nucleon in terms of $m_\psi$ for two different values of Higgs mixing angle; $\theta=0.01$ and $\theta=0.1$.  }\label{fig1.}
\end{figure}
where $m_\text{r} =\left(\frac{1}{m_{\psi}}+\frac{1}{m_{\text{p}}}\right)^{-1}$.

 Using $g_s$ as obtained in the previous subsection for $\theta=0.1$ and $0.01$ we plot the direct detection cross section in Fig. \ref{fig1.}. We also compare our result with the new updated experimental data in this figure.  The data which we have used here are from the XENON100 \cite{XENON100}, PandaX II \cite{pandax}, LUX \cite{LUX2016},  PICO-60 \cite{pico60}  and Darkside-50 \cite{darkside} Collaborations.

\section{Discussion and conclusions}
The most minimal and renormalizable extension of the SM, which introduces a singlet fermion as CDM candidate, is the SFCDM model. Namely, one adds a singlet fermion as CDM and a scalar as mediator to the SM content. A comprehensive analysis of this model has been given in \cite{ettefaghi}. However, the mixing angle between the SM Higgs and singlet scalar is constrained to be less than 0.1 \cite{adam}. Therefore, we have restudied the relevant parameter space to determine the role of the mixing angle. The SM Higgs boson mass is fixed to 125 GeV according to the 2012 ATLAS \cite{atlas} and CMS \cite{cms} reports and the other Higgs mass to 750 GeV as we have explained in the main body of paper. In order to find the coupling $g_s$ which satisfies the relic density condition, we first investigate about 25000 sample models randomly in the whole parameter space. In fact, in addition to $\lambda$, we take $\theta$ and $m_\psi$ to be free. The data of this study is denoted by blue points in Fig. \ref{gs}. We see that $g_s$ tends to a unique value for $m_\psi$ larger than about 750 GeV. Two other investigations with fixed $\theta=0.1$ and $\theta=0.01$, each of which with 10000 sample models, have been denoted in Fig. \ref{gs} by orange and green points, respectively.
For more clarification, we illustrate the behavior of $g_s$ in terms of $m_\psi$ and $\theta$ through Fig. \ref{theta}. We see that there exist limited regions (300 GeV$<m_\psi<400$ GeV and 700 GeV$<m_\psi<1000$ GeV) in which $\theta<0.1$ and $g_s<1$. Furthermore, after deriving the spin-independent cross section of the elastic scattering of SFCDM from nucleon, we use the $g_s$  obtained from relic abundance condition to calculate and plot this cross section. It is illustrated through Fig. \ref{fig1.} in terms
of $m_\psi$ for two various choices of $\theta$. We have compared our results with different experimental data. According to this figure, the entire parameter space is excluded by XENON100 \cite{XENON100}, LUX \cite{LUX2016} and PandaX II  \cite{pandax}. For more comparison, we have also shown the recent experiments PICO-60 \cite{pico60} and DarkSide-50 \cite{darkside}  in this figure.


\begin{thebibliography}{99}
	\bibitem{rev1}
	G. Jungman, M. Kamionkowski, and K. Griets, ``Supersymmetric dark matter" \textit{Phys. Rep.} {\bf 267}  (1996) 195.
	
	\bibitem{rev2}
	G. Bertone, D. Hooper, and J. Silk, ``Particle dark matter: Evidence, candidates and constraints", \textit{Phys. Rep.} {\bf 405} (2005) 279.
	
	\bibitem{Taoso}
	M.Taoso , G.Bertone, and A.Masiero, ``Dark matter candidates: a ten-point test", \textit{JCAP} {\bf 0803}  (2008) 022.
	
	\bibitem{kim1}
	Y. G. Kim, and K. Y. Lee, ``Minimal model of fermionic dark matter", \textit{Phys. Rev. D} \textbf{75} (2007) 115012.
	
	\bibitem{kim2}
	Y. G. Kim, K. Y. Lee, and S. Shin, ``Singlet fermionic dark matter", \textit{JHEP} \textbf{0805} (2008) 100.
	
	\bibitem{Beak}
	S. Baek, P. Ko, W.-I. Park, and E. Senaha, ``Search for the Higgs portal to a singlet fermionic dark matter at the LHC",  \textit{JHEP} \textbf{1211} (2012) 116.
	
	\bibitem{ettefaghi}
	Z. Bagherian, M.M. Ettefaghi, Z. Haghgouyan and R. Moazzemi, ``A new parameter space study of the fermionic cold dark matter model", \textit{JCAP} \textbf{10} (2014) 033.
	
	\bibitem{ettefaghi2}
	M.M. Ettefaghi and R. Moazzemi, ``Annihilation of singlet fermionic dark matter into two photons", \textit{JCAP} \textbf{02} (2013) 048 [arXiv:1301.4892].
	
	\bibitem{atlastheta}
	ATLAS and CMS Collaboration, ``Measurements of the Higgs boson production and decay
	rates and constraints on its couplings from a combined ATLAS and CMS analysis of the LHC \textit{pp} collision data at $\sqrt{s}=7$ and 8 TeV", \textit{JHEP} \textbf{08} (2016) 045.
	
	\bibitem{falkow}
	A. Falkowski, C. Gross, and O. Lebedev, ``A Second Higgs from the Higgs Portal", \textit{JHEP} \textbf{05}
	(2015) 057, [arXiv:1502.01361].
	
	\bibitem{adam}
	A. Falkowski, O. Sloneb and Tomer Volansky, ``Phenomenology of a 750 GeV singlet", \textit{JHEP} \textbf{02} (2016) 152 [arXiv:1512.05777v3].
	
	\bibitem{atlas2015}
	ATLAS Collaboration, M. Aaboud, et al.``Search for resonances in diphoton events at $\sqrt{s}=$ 13 TeV with the ATLAS detector", \textit{JHEP} \textbf{09} (2016) 001 [arXiv:1606.03833].
	
	\bibitem{cms2015}
	CMS Collaboration, V. Khachatryan, et al.``Search for resonant production of high-mass photon pairs in
	proton-proton collisions at $\sqrt{s}$ = 8 and 13 TeV", \textit{Phys. Rev. Lett.} \textbf{117} (2016) 051802 [arXiv:1606.04093].
	
\bibitem{atlas2016}
	ATLAS collaboration, ``Search for scalar diphoton resonances with 15.4 fb− 1 of data collected at $\sqrt{s}= 13$ TeV in 2015 and 2016 with the ATLAS detector." Tech. Repor t ATLAS-CONF-2016. 2016.
	
\bibitem{cms2016}
CMS collaboration, ``Search for resonant production of high mass photon pairs using 12.9 fb$^{− 1}$ of proton-proton collisions at $\sqrt{s}= 13$ TeV and combined interpretation of searches at 8 and 13 TeV." CMS PAS EXO-16-027 (2016).
	
	\bibitem{pico-2l}
	PICO Collaboration, C. Amole, et al. ``Improved dark matter search results from PICO-2L Run 2" \textit{Phys. Rev. D} \textbf{93} (2016) 061101.
	
	
	\bibitem{gondolo}
	P. Gondolo and G. Gelmini, ``Cosmic abundances of stable particles: improved analysis",
	\textit{Nucl. Phys. B} 360 (1991) 145.
	
	\bibitem{colb}
	E.W. Kolb and M.S. Turner, The early universe, (Addison Wesley, New York 1990).
	
	
	\bibitem{Spergel}
	WMAP Collaboration, D.N. Spergel et al, ``Wilkinson microwave anisotropy probe (wmap)
	three year results: Implications for cosmology", \textit{Astrophys. J. Suppl. }{\bf 170} (2007) 377 [astro-ph/0603449].
	
	\bibitem{atlas}
	ATLAS Collaboration, G. Aad et al., ``Observation of a new particle in the search for the standard model Higgs boson with the ATLAS detector at the LHC", \textit{Phys. Lett.} \textbf{B716}  (2012) 1 [arXiv:1207.7214].
	

	
	\bibitem{cms}
	CMS Collaboration, S. Chatrchyan et al., ``Observation of a new boson at a mass of 125 GeV with the CMS experiment at the LHC", \textit{Phys. Lett.} \textbf{B716} (2012) 30 [arXiv:1207.7235].
	
\bibitem{XENON100}
			XENON100 Collaboration, E. Aprile, et al. ``XENON100 dark matter results from a combination of 477 live days." \textit{Phys. Rev. D} \textbf{94} (2016) 122001.
			
				\bibitem{pandax}
				PandaX-II Collaboration, A. Tan,  et al. ``Dark matter results from first 98.7-days of data from the PandaX-II experiment" \textit{Phys. Rev. Lett. } \textbf{117} (2016) 121303.
			
\bibitem{LUX2016}
			LUX Collaboration, D.S. Akerib, et al. ``Results from a search for dark matter in LUX with 332 live days of exposure",\textit{ Phys. Rev. Lett.} \textbf{118} (2017) 021303.
	
	\bibitem{Ellis}
	J. Ellis, A. Ferstl and K.A. Olive, ``Re-evaluation of the elastic scattering of supersymmetric dark matter",\textit{ Phys. Lett.} \textbf{B481} (2000) 304.
	
	\bibitem{pico60}
	PICO Collaboration, C. Amole, et al. ``Dark matter search results from the PICO-60 CF 3 I bubble chamber", \textit{Phys. Rev. D} \textbf{93} (2016) 052014.
	
	

	
	\bibitem{darkside}
	DarkSide Collaboration, P. Agnes, et al. ``Results from the first use of low radioactivity argon in a dark matter search" \textit{Phys. Rev. D } \textbf{93} (2016) 081101.
	
\end{thebibliography}
\end{document}